\documentclass[onecolumn,preprintnumbers,notitlepage,superscriptaddress,
nofootinbib,amsfont,amssymb,10pt,singlespacing] {revtex4-1}
\usepackage{amsmath,bm}
\usepackage{times}
\usepackage{braket}
\usepackage{color,graphicx}
\usepackage{slashed}
\usepackage{hyperref}
\usepackage{graphics}
\usepackage{subfigure}
\usepackage{multirow,makecell}
\usepackage{textcomp}
\usepackage{mathtools}
\usepackage{float}

\newcommand{\beq}{\begin{eqnarray}}
\newcommand{\eeq}{\end{eqnarray}}

\def\lsim{ {\ \lower-1.2pt\vbox{\hbox{\rlap{$<$}\lower6pt\vbox{\hbox{$\sim$}
}}}\ } }
\def\gsim{ {\ \lower-1.2pt\vbox{\hbox{\rlap{$>$}\lower6pt\vbox{\hbox{$\sim$}
}}}\ } }
\definecolor{Red}{rgb}{1.,0.,0.}

\definecolor{Blue}{rgb}{0.,0.,1.}

\definecolor{nicered}{rgb}{0.7,0.1,0.1}
\definecolor{nicegreen}{rgb}{0.1,0.5,0.1}

\hypersetup{colorlinks,citecolor=nicegreen,linkcolor=nicered}

\begin{document}

\title{
Study of the quasi-two-body decays $B^{0}_{s} \rightarrow \psi(3770)(\psi(3686))\pi^+\pi^-$ with perturbative QCD approach}
%%%==================================================================

\author{Ze-Rui~Liang}
\email[Electronic address:]{Liangzr@email.swu.edu.cn}
\affiliation{School of Physical Science and Technology,
 Southwest University, Chongqing 400715, China}

\author{Feng-Bo~Duan}
\email[Electronic address:]{dfbdfbok@163.com}
\affiliation{School of Physical Science and Technology,
 Southwest University, Chongqing 400715, China}

\author{Xian-Qiao~Yu}
\email[Electronic address:]{yuxq@swu.edu.cn}
\affiliation{School of Physical Science and Technology,
Southwest University, Chongqing 400715, China}

\date{\today}

%%%%%%%%%%%%%%%%%%%%%%%%%%%%%%%%%%%%%%%%%%%%%%%%%%%%%%%%%%%%%%%%%%
\begin{abstract}

In this note, we study the contributions from the $S$-wave resonances, $f_{0}(980)$ and $f_{0}(1500)$, to the $B^{0}_{s}\rightarrow \psi(3770)\pi^ {+}\pi^{-}$ decay by introducing the $S$-wave $\pi\pi$ distribution amplitudes within the framework of the perturbative QCD approach. Both resonant and nonresonant contributions are contained in the scalar form factor in the $S$-wave distribution amplitude $\Phi^S_{\pi\pi}$. Since the vector charmonium meson $\psi(3770)$ is a $S-D$ wave mixed state, we calculated the branching ratios of $S$-wave and $D$-wave respectively, and the results indicate that $f_{0}(980)$ is the main contribution of the considered decay, and the branching ratio of the $\psi(2S)$ mode is in good agreement with the experimental data. We also take the $S-D$ mixed effect into the $B^{0}_{s}\rightarrow \psi(3686)\pi^ {+}\pi^{-}$ decay. Our calculations show that the branching ratio of $B^{0}_{s}\rightarrow \psi(3770)(\psi(3686))\pi^ {+}\pi^{-}$ can be at the order of $10^{-5}$, which can be tested by the running LHC-b experiments.

\end{abstract}
%%%%%%%%%%%%%%%%%%%%%%%%%%%%%%%%%%%%%%%%%%%%%%%%%%%%%%%%%%%%%%

\maketitle

%
%%%
%%%%%%%%%%%%%%%%% I. INTRODUCTION %%%%%%%%%%%%%%%%%%%%%%%%%%%%%%%%
%%%
%

\section{Introduction}\label{sec:intro}

In this decade, the $B$ mesons' three-body hadronic decays have drawn lots of attention on both experimental and theoretical sides, since it can test the standard model(SM) and help us to have a better understanding of the scheme of QCD dynamics. The three-body decays of $B$ mesons are more complicated than the two-body cases because it include both resonant and nonresonant contributions and have two-gluon exchange in the decay amplitude within our framework of theoretical calculation, which will lead to the hard kernel that involves 3-body, and may introduce the possible final-state interactions~\cite{Bediaga:2013ela,Bediaga:2015mia,Magalhaes:2017yov,Kang:2013jaa}. So just as stated in Ref.~\cite{Cheng:2007si}, because of the interference between the resonance and nonresonance, it is difficult to make a direct calculation of the resonance and non-resonance contributions separately.

In experiment, the Heavy Flavor Averaging Group~\cite{Amhis:2016xyh} have collected lots of world averages of measurements of B-hadron properties from LHCb~\cite{Aaij:2013cpa,Aaij:2014vda,LHCb:2012ae,Aaij:2014siy,Aaij:2014emv,Aaij:2017zpx,Aaij:2014dka,Aaij:2017ngy}, Belle~\cite{Garmash:2004wa} and BaBar~\cite{Lees:2015uun,Aubert:2002vb,BABAR:2011aaa} and other collaborations, in which the LHCb Collaboration have measured sizable direct CP asymmetries in kinematic regions.
 Due to the progress of the experiments, three-body decays have been analyzed by different methods, which based on the symmetry principles and factorization theorems, however, the theoretical studies is still at an early stage. Many authors have studied the three-body decays with the improved QCD factorization~\cite{Zhang:2013oqa,Krankl:2015fha,Klein:2017xti,Furman:2005xp,ElBennich:2006yi,ElBennich:2009da,Wang:2016yrm,Lu:2010vb,Cheng:2002qu} and the perturbative QCD (pQCD) factorization approachs~\cite{Lu:2018fqe,Li:2016tpn,Wang:2017hao,Li:2017mao,Li:2018lbd,Rui:2018hls,Wang:2015uea,Chen:2004az,Chen:2002th,Wang:2014ira,Wang:2016rlo,Rui:2017bgg}. The inside nonresonance contributions have been performed by using heavy meson chiral perturbative theory approach in Refs.~\cite{Cheng:2007si,Cheng:2002qu,Cheng:2013dua} and the references therein, and the resonance contributions are treated using the isobar model~\cite{Herndon:1973yn} within the Breit-Wigner formalism~\cite{Breit:1936zzb}.

In the pQCD approach which is based on the $\emph{k}_{T}$ factorization theorem, the three-body decay can be simplified into two-body case by bringing in two-hadron distribution amplitudes~\cite{Diehl:1998dk,Diehl:2000uv}, which contain the messages of both resonance and nonresonance.
The dominant contributions come from the region, where the two light meson pair moves parallelly with an invariant mass below $\emph{O}(\bar{\Lambda}M_{B})$, where $\bar{\Lambda}=M_{B}-m_{b}$ is the mass difference between the $B$ meson and $b$-quark. Therefore, one can express the typical pQCD factorization formula of the $B$ meson's three-body decay amplitude as~\cite{Chen:2004az,Chen:2002th}

\begin{equation}
{\cal A}={\cal H}\otimes \phi_{B} \otimes \phi_{h_{3}} \otimes \phi_{h_{1}h_{2}},
\end{equation}

where the hard decay kernel ${\cal H}$ describes the contribution from the region with only one gluon exchange diagram at leading order, that can be calculated in the perturbative theory. The $\phi_{B}$, $\phi_{h_{1}h_{2}}$ and $\phi_{h_{3}}$, which can be extracted from experiment or calculated by several nonperturbative approach and can be regarded as nonperturbative input, is the distribution amplitude of B meson, $h_1h_2$ pair and $h_3$, respectively.

The $B^{0}_{s} \rightarrow \psi(2S)\pi^{+}\pi^{-}$ decay was first observed by the LHCb collaboration~\cite{Aaij:2013cpa}, the data was displayed based on an integrated luminosity of 1 $fb^{-1}$ in the pp collisions at a centre-of-mass energy of $\sqrt{s}=7$ {\rm TeV}. It is found that $f_{0}(980)$ is the main source of the decay rate by a way called $\emph{sPlot}$ technique. The $B^{0}_{s} \rightarrow \psi(3770)\pi^{+}\pi^{-}$ decay have not been observed yet, so it is desirable to make a theoretical prediction for the branching ratios of this decay mode, for testing the three body decay's mechanism and the mixing scheme of $\psi(3770)$ as well. In this work, we will calculate the branching ratio of the quasi-two-body decay mode $B^{0}_{s} \rightarrow \psi(3770) \pi^+\pi^-$, since the vector charmonium meson $\psi(3770)$, the lowest-lying charmonium state just above the $D\bar{D}$ threshold, is mainly regarded as a $S-D$ mixture, we will take the $S$-wave and $D$-wave contribution into account respectively. Here, $\psi(2S)$ is the first radially excited charmonium meson, and the pure $1D$ state indicates the principle quantum number $n=1$ and the orbital quantum number $l=2$. The $S-D$ mixing angle $\theta$ can be obtained from the ratio of the leptonic decay widths of $\psi(3686)$ and $\psi(3770)$~\cite{Eidelman:2004wy}. In Ref.~\cite{Gao:2006yu}, the authors make tentative calculations for different mixing solutions of the $B$ meson exclusive decay $B \rightarrow \psi(3770)K$ with the QCD factorization, and they drew a conclusion that when taking account of higher-twist effects and adopting the $S-D$ mixing angle $\theta=-(12\pm2)^{\circ}$, the widely accepted value, the branching ratio of the decay $B \rightarrow \psi(3770)K$ can fit the experimental data well. Also, in Refs.~\cite{Rosner:2001nm,Kuang:2002hz,Ding:1991vu}, the authors provided two sets of mixing scheme within the nonrelativistic potential model: $\theta=-(12\pm2)^{\circ}$ or $\theta=(27\pm2)^{\circ}$. Here, the charmonium mesons $\psi(3686)$ and $\psi(3770)$ may be almost described as~\cite{Rosner:2001nm,Kuang:2002hz,Ding:1991vu,Duan:2017rhr}

\begin{equation}
\begin{split}
\psi(3686)=\cos\theta|c\bar{c}(1D)\rangle+\sin\theta|c\bar{c}(2S)\rangle,\\
\\
\psi(3770)=\cos\theta|c\bar{c}(1D)\rangle-\sin\theta|c\bar{c}(2S)\rangle.
\end{split}
\end{equation}

The content syllabus of this paper is as follows. After the introduction, we describe the theoretical framework and the wave function of the excited charmonium mesons $\psi(2S)$ and $\psi(1D)$ in section \ref{sec:pert}. And in section \ref{sec:amp}, we list the decay amplitude of the considered decay modes. The numerical results and analysis about the results we have got will be shown in section \ref{sec:numer}. Finally we will finish this paper with a brief summary.

%
%%%
%%%%%%%%%%%%%%%%% II. Theoretical frame and the wave function %%%%%%%%%%%%%%%%%%%%%%%%%%%%%%%%
%%%
%
\section{The Theoretical Framework And The Wave Function}\label{sec:pert}

For the quasi-two-body $B^{0}_{s} \rightarrow \psi f_{0}(\rightarrow\pi^{+}\pi^{-})$ decays, the relevant weak effective Hamiltonian can be written as~\cite{Buchalla:1995vs}

\begin{equation}
{\cal H}_{eff}=\frac{G_{F}}{\sqrt2}\big\{V^*_{cb}V_{cs}[C_{1}(\mu)O_{1}(\mu)+C_{2}(\mu)O_{2}(\mu)]-V^*_{tb}V_{ts}[\sum^{10}_{i=3}C_{i}(\mu)O_{i}(\mu)]\big\}+H.c.,
\end{equation}

where $V^*_{cb}$$V_{cs}$ and $V^{*}_{tb}$$V_{ts}$ are Cabibbo-Kobayashi-Maskawa (CKM) factors, $O_{i}(\mu)$ is local four-quark operator and $C_{i}(\mu)$ is corresponding Wilson coefficient.

\begin{figure}[htbp]
 \centering
 \begin{tabular}{l}
 \includegraphics[width=0.5\textwidth]{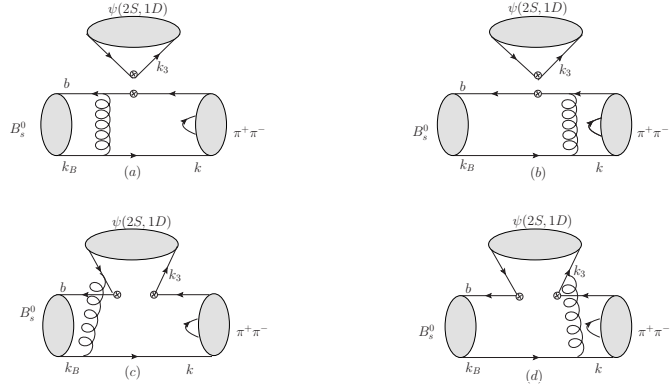}
 \end{tabular}
 \caption{The lowest order Feynman diagrams for the $B^{0}_{s} \rightarrow \psi f_{0}(\rightarrow\pi^{+}\pi^{-})$ decays}
   \label{fig:figure1}
 \end{figure}

It is convenient for us to choose the light-cone coordinates for simplicity. In this coordinates, we choose the $B^0_{s}$ meson at rest, and let the $\pi\pi$ meson pair and $\psi(2S,1D)$ meson move along with the direction of $n=(1,0,0_{\top})$ and $v=(0,1,0_{\top})$, respectively, and the Feynman diagrams have been described in Fig.~\ref{fig:figure1}. So the momentums of the $B^0_{s}$ ($p_{B}$), $\pi\pi$ ($p$), and $\psi(2S,1D)$ ($p_{3}$) are written as

\begin{equation}
\begin{split}
p_{B}=\frac{M_{B^{0}_{s}}}{\sqrt{2}}(1,1,0_{\top}),                 \\
p=\frac{M_{B^{0}_{s}}}{\sqrt{2}}(1-r^2,\eta,0_{\top}),              \\
p_{3}=\frac{M_{B^{0}_{s}}}{\sqrt{2}}(r^2,1-\eta,0_{\top}).
\end{split}
\end{equation}

Meanwhile, the corresponding light quark's momentum in each meson read as
\begin{equation}
\begin{split}
k_{B}=(0,\frac{M_{B^{0}_{s}}}{\sqrt{2}}{\emph{x}_{B}},k_{B\top}),\\
k=(\frac{M_{B^{0}_{s}}}{\sqrt{2}}\emph{z}(1-r^2),0,k_{\top}),\\
k_{3}=(\frac{M_{B^{0}_{s}}}{\sqrt{2}}r^2\emph{x}_{3},\frac{M_{B^{0}_{s}}}{\sqrt{2}}(1-\eta)\emph{x}_{3},k_{3\top}),
\end{split}
\end{equation}

here, $M_{B^{0}_{s}}$ is the mass of $B^{0}_{s}$, and $r=\frac{M_\psi}{M_{B^{0}_{s}}}$ is the corresponding mass ratio, $M_{\psi}$ denotes the $\psi(2S,1D)$ mesons mass. The variable $\eta=\omega^2/({M_{B^{0}_{s}}^2}-{M_\psi}^2)$, where the pion-pair invariant mass $\omega^2$ and its momentum $p$ satisfy the relation $\omega^2=p^2$ and $p=p_{1}+p_{2}$. $\emph{x}_{1}$, $\emph{z}$, and $\emph{x}_{3}$ indicate momentum fractions of the spectator quark inside the meson, they are in the range of $0\sim1$. By introducing the kinematic variables, $\zeta$, of the pion-pair, we define $\zeta=p^+_{1}/p^+$ as the $\pi^+$ meson momentum fraction, the other component's kinematic variables of pion-pair can be expressed as
\begin{equation}
\begin{split}
p^-_{1}=\frac{M_{B^{0}_{s}}}{\sqrt{2}}(1-\zeta)\eta ,  \\                          p^+_{2}=\frac{M_{B^{0}_{s}}}{\sqrt{2}}(1-\zeta)(1-r^2),\\
p^-_{2}=\frac{M_{B^{0}_{s}}}{\sqrt{2}}\zeta\eta .
\end{split}
\end{equation}
In our calculations, the hadron $B^{0}_{s}$ usually treated as a heavy-light system, the wave function of which can be found in Refs.~\cite{Lu:2000em,Keum:2000wi,Keum:2000ph}
\begin{equation}
\Phi_{B^{0}_{s}}=\frac{i}{\sqrt{2\emph{N}_{c}}}(\not {p}_{B}+M_{B^{0}_{s}}){\gamma_{5}}{\phi_{B_{s}}({\emph{x}_{B},\emph{b}_{B}})},
\end{equation}
where the distribution amplitude(DA) ${\phi_{B_{s}}({\emph{x}_{B},\emph{b}_{B}})}$ of $B^{0}_{s}$ meson is written as mostly used form, which is
\begin{equation}
\phi_{B_{s}}(\emph{x}_{B},\emph{b}_{B})=\emph{N}_{B}{{\emph{x}_{B}}^2}(1-{\emph{x}_{B}})^2\exp[-\frac{M^2_{B^{0}_{s}}{{\emph{x}_{B}}^2}}{2\omega^2_{B_{s}}}-\frac{1}{2}(\omega_{B_{s}}{\emph{b}_{B}})^2],
\end{equation}
the normalization factor $\emph{N}_{B}$ can be calculated by the normalization relation $\int^{1}_{0}\emph{dx}\phi_{B_{s}}(\emph{x}_{B},\emph{b}_{B}=0)=\emph{f}_{B^{0}_{s}}/({2}{\sqrt{{2}{\emph{N}}_{c}}})$ with $\emph{N}_{c}=3$ is the color number. Here, we choose shape parameter $\omega_{B_{s}}=0.50\pm0.05$ {\rm GeV}~\cite{Ali:2007ff}.

For the vector charmonium meson $\psi(3770)$, as mentioned above, it is commonly regarded as a $S$-wave and $D$-wave mixing state. We adopt the wave function form of this vector charmonium meson with the basis of harmonic-oscillator potential, which have been applied to the charmonium state successfully, such as $J/\psi$, $\psi(2S)$, $\psi(3S)$ and so on~\cite{Yu:2009zq,Rui:2015iia,Rui:2016opu,Duan:2017rhr}, and the theoretical results agree well with the measured experimental data, which indicate the reasonability to adopt this form of the function. For the wave function of the pure $2S$ state, $\psi(2S)$, and the pure $1D$ state, $\psi(1D)$, whose longitudinal polarized component is defined as~\cite{Rui:2015iia,Rui:2016opu}

\begin{equation}
\Phi^{L}_{\psi}=\frac{1}{\sqrt{2\emph{N}_{c}}}[M_{\psi}\not{\epsilon}_{L}\psi^L(\emph{x}_{3},\emph{b}_{3})+\not{\epsilon}_{L}\not {p}_{3}\psi^t(\emph{x}_{3},\emph{b}_{3})]
\end{equation}

where $p_{3}$ is the momentum of the charmonium mesons $\psi(2S)$, $\psi(1D)$ with the longitudinal polarization vector $\epsilon_{L}=\frac{M_{B^0_{s}}}{\sqrt{2} M_{\psi}}(-{r^2},(1-\eta),0_{\top})$ and $M_{\psi}$ is the corresponding mass.
Here the $\psi^{\emph{L}}$ and $\psi^{\emph{t}}$ attribute to twist-2 and twist-3 distribution amplitudes(DAs). The explicit form are:~\cite{Rui:2015iia,Duan:2017rhr}
\begin{equation}
\psi^{\emph{L}}(\emph{x}_{3},\emph{b}_{3})= \frac {\emph{f}_{(2S,1D)}}{2\sqrt{2\emph{N}_{c}}}N^{\emph{L}}\emph{x}_{3}\overline{\emph{x}}_{3} {\cal I}(\emph{x}_{3})e^{-\emph{x}_{3}\overline{\emph{x}}_{3}\frac{\emph{m}_c}{\omega}
[\omega^2\emph{b}_{3}^2+(\frac{\emph{x}_{3}-\overline{\emph{x}}_{3}}{2\emph{x}_{3}\overline{\emph{x}}_{3}})^2]},
\label{eq:exp10}
\end{equation}

\begin{equation}
\psi^{\emph{t}}(\emph{x}_{3},\emph{b}_{3})= \frac {\emph{f}_{(2S,1D)}}{2\sqrt{2N_{c}}}N^{\emph{t}} ({\emph{x}_{3}-\overline{\emph{x}}_{3}})^2{\cal I}(\emph{x}_{3})e^{-\emph{x}_{3}\overline{\emph{x}}_{3}\frac{\emph{m}_c}{\omega}
[\omega^2\emph{b}_{3}^2+(\frac{\emph{x}_{3}-\overline{\emph{x}}_{3}}{2\emph{x}_{3}\overline{\emph{x}}_{3}})^2]},
\label{eq:exp11}
\end{equation}
with ${\cal I}(\emph{x}_{3})=1-4{\emph{m}_c}\omega {\emph{x}_{3}\overline{\emph{x}}_{3}}\emph{b}_{3}^2+\frac{\emph{m}_c(1-2\emph{x}_{3})^2}{\omega \emph{x}_{3}\overline{\emph{x}}_{3}}$ for $\psi(2S)$ and ${\cal I}(\emph{x}_{3})=(\frac{1}{\emph{x}_{3}\overline{\emph{x}}_{3}}-\emph{m}_{c}\omega \emph{b}_{3}^2)(6\emph{x}_{3}^4-12\emph{x}_{3}^3+7 \emph{x}_{3}^2-\emph{x}_{3})-\frac{\emph{m}_c(1-2\emph{x}_{3})^2}{4\omega \emph{x}_{3}\overline{\emph{x}}_{3}}$ for $\psi(1D)$.
The shape parameter $\omega_{1D}$ in the DAs of the $\psi(1D)$, we choose $\omega_{1D}=0.5\pm0.05$ {\rm GeV}, for the reason we have discussed in Ref.~\cite{Duan:2017rhr}, and $\omega_{2S}=0.2\pm0.1$ {\rm GeV}~\cite{Rui:2015iia}. The $N^{\emph{i}}(i=L,t)$ is the normalization constant, which satisfy the normalization conditions:

\begin{equation}
\int^1_0 \psi^{\emph{i}}(\emph{x}_{3},\emph{b}_{3}=0)d{\emph{x}_{3}}= \frac {\emph{f}_{(2S,1D)}}{2\sqrt{2\emph{N}_{c}}},
\end{equation}
and the decay constant of the radially excited state $\psi(2S)$ and the angular excitation state $\psi(1D)$ have been given in Table \ref{tab}.
Both the wave functions Eq.~(\ref{eq:exp10}) and Eq.~(\ref{eq:exp11}) are symmetric under $\emph{x}\leftrightarrow \overline{\emph{x}}$.

In the light of Ref.~\cite{Meissner:2013hya,Doring:2013wka}, we adopt the distribution amplitudes for $S$-wave pion-pair as
\begin{equation}
\Phi^{S}_{\pi\pi}=\frac{1}{\sqrt{2 \emph{N}_{c}}}[\not {p}\phi^{I=0}_{v\nu=-}(\emph{z},\zeta,\omega^2)+\omega\phi^{I=0}_{s}(\emph{z},\zeta,\omega^2)+\omega(\not{n}\not {v}-1)\phi^{I=0}_{t\nu=+}(\emph{z},\zeta,\omega^2)].
\end{equation}

For simplicity, we put $\phi^{I=0}_{v\nu=-}(\emph{z},\zeta,\omega^2)$, $\phi^{I=0}_{s}(\emph{z},\zeta,\omega^2)$ and $\phi^{I=0}_{t\nu=+}(\emph{z},\zeta,\omega^2)$ abbreviated to $\phi_{0}$, $\phi_{s}$, and $\phi_{\sigma}$, respectively. The relevant DAs and time-like scalar form factor can be get from Ref.~\cite{Wang:2015uea, Wang:2015paa, Shi:2015kha}.

The differential branching ratios for the $B^{0}_{s} \rightarrow \psi(2S,1D)\pi^+\pi^-$ decay in the $B^{0}_{s}$ meson rest frame can be written as~\cite{Tanabashi:2018oca}
\begin{equation}\label{equ:dbr}
\frac{d\cal{B}}{d\omega}=\frac{\tau_{B^0_{s}}\omega \mid {\overrightarrow{p_{1}}} \mid \mid {\overrightarrow{p_{3}}} \mid}{32(\pi M_{B^0_{s}})^3} \mid {\cal A} \mid^2,
\end{equation}
with $p_{1}=\frac{1}{2} \sqrt{\omega^2-4 m^2_{\pi^{\pm}}}$ and $p_{3}=\frac{1}{2\omega}\sqrt{[M^2_{B^0_{s}}-(\omega+M_{\psi})^2][M^2_{B^0_{s}}-(\omega-M_{\psi})^2]}$ in the pion-pair center-of-mass system and the $B^0_{s}$ meson lifetime $\tau_{B^0_{s}}$.

\section{The decay Amplitudes}\label{sec:amp}
In the pQCD factorization apporach, the $B^{0}_{s} \rightarrow \psi(2S)\pi^+\pi^-$ decay amplitude $\cal{A}$ express as in form of
\begin{equation}
{\cal{A}}=V^*_{cb}V_{cs}(\emph{F}^{(V-A)(V-A)}+\emph{M}^{(V-A)(V-A)})-V^*_{tb}V_{ts}(\emph{F}^{'(V-A)(V-A)}+\emph{F}^{(V-A)(V+A)}+\emph{M}^{'(V-A)(V-A)}+\emph{M}^{(S-P)(S+P)}),
\label{eq:amp}
\end{equation}
where the explicit form of $\emph{F}^{(V-A)(V-A)}$, $\emph{F}^{'(V-A)(V-A)}$, $\emph{F}^{(V-A)(V+A)}$, and $\emph{M}^{(V-A)(V-A)}$, $\emph{M}^{'(V-A)(V-A)}$, $\emph{M}^{(S-P)(S+P)}$ are listed as follows and $\emph{F}$, $\emph{M}$ denote the factorization and non-factorization contribution respectively.
$(V-A)(V-A)$, $(V-A)(V+A)$ are the weak vertexes of the operators, and $(S-P)(S+P)$ denotes the Fierz transformation of the $(V-A)(V+A)$.
\begin{equation}
\begin{split}
{\emph F}^{(V-A)(V-A)}=&8\pi\emph{C}_{\emph{F}}f_{\psi}M^{4}_{B^0_{s}}\int^{1}_{0}\emph{d}\emph{x}_{B}\emph{d}\emph{z}\int^\infty_{0}\emph{b}_{B}\emph{b}\emph{d}\emph{b}_{B}\emph{d}\emph{b}\phi_{B_{s}}(\emph{x}_{B},b_{B})\\
& \times \{[(\overline{\eta} (1+\emph{z}(1-2r^2))-r^2) \phi_{0}+ \sqrt{\eta(1-r^2)}[((1-2\emph{z}(1-r^2)) \overline{\eta}-r^2)( \phi_{s}+ \phi_{\sigma})+2r^2 \phi_{\sigma}]]\\
& \times \alpha_{s}(t_{a})\emph{a}_{1}(t_{a})h_{a}(\emph{x}_{B}, \emph{z}, \emph{b}_{B}, \emph{b})S_{t}(\emph{z})\exp[-S_{B^{0}_{s}}(t_{a})-S_{M}(t_{a})]\\
& +[(r^2-1)[\overline{\eta}\eta+r^2(\emph{x}_{B}-\eta)]\phi_{0}+2\sqrt{\eta(1-r^2)}[\overline{\eta}-r^2(1-\emph{x}_{B})]\phi_{s}]\\
& \times \alpha_{s}(t_{b})\emph{a}_{1}(t_{b})h_{b}(\emph{x}_{B}, \emph{z}, \emph{b}_{B}, \emph{b})S_{t}(|\emph{x}_{B}-\eta|)\exp[-S_{B^{0}_{s}}(t_{b})-S_{M}(t_{b})]\},
\end{split}
\end{equation}
\begin{equation}
\begin{split}
{\emph{F}}^{'(V-A)(V-A)}={\emph{F}}^{(V-A)(V-A)}|_{\emph{a}_{1}\rightarrow \emph{a}_{2}},
\end{split}
\end{equation}
\begin{equation}
\begin{split}
{\emph{F}}^{(V-A)(V+A)}={\emph{F}}^{(V-A)(V-A)}|_{\emph{a}_{1}\rightarrow \emph{a}_{3}},
\end{split}
\end{equation}
\begin{equation}
\begin{split}
{\emph{M}}^{(V-A)(V-A)}=&\frac{-32\pi\emph{C}_{\emph{F}}M^{4}_{B^{0}_{s}}}{\sqrt{2\emph{N}_{c}}}\int^{1}_{0}\emph{d}\emph{x}_{B}\emph{d}\emph{z}\emph{d}\emph{x}_{3}\int^\infty_{0}\emph{b}_{B}\emph{b}_{3}\emph{d}\emph{b}_{B}\emph{d}\emph{b}_{3}\phi_{B_{s}}(\emph{x}_{B},b_{B})\\
& \times \{[(\overline{\eta}-r^2)[((1-\emph{x}_{3}-\emph{x}_{B})(1-r^2)+\eta(\emph{x}_{3}(1-2r^2)\\
&-(1-r^2)(1-\emph{z})+r^2))\psi^{L}(\emph{x}_{3}, \emph{b}_{3})+r r_{c}\overline{\eta}\psi^{t}(\emph{x}_{3}, \emph{b}_{3})]\phi_{0}\\
& +\sqrt{\eta(1-r^2)}[\overline{\eta}(\emph{z}(1-r^2)+2r^2(1-\emph{x}_{3})-r^2\emph{x}_{B})\phi_{\sigma}-(\emph{z}\overline{\eta}(1-r^2)+r^2\emph{x}_{B})\phi_{s}]\psi^{L}(\emph{x}_{3}, \emph{b}_{3})]\\
& \times \alpha_{s}(t_{c})\emph{C}_{2}(t_{c})h_{c}(\emph{x}_{B}, \emph{z}, \emph{x}_{3}, \emph{b}_{B}, \emph{b}_{3})\exp[-S_{B^{0}_{s}}(t_{c})-S_{M}(t_{c})-S_{\psi}(t_{c})]\\
& +[(\overline{\eta}-r^2)[(\emph{x}_{B}-\emph{z}(1-r^2)-\emph{x}_{3}(\overline{\eta}+r^2))\psi^{L}(\emph{x}_{3}, \emph{b}_{3})+r r_{c} \overline{\eta}\psi^{t}(\emph{x}_{3}, \emph{b}_{3})]\phi_{0}\\
& -\sqrt{\eta(1-r^2)}[((r^2\emph{x}_{B}-\overline{\eta}(2r^2 \emph{x}_{3}+\emph{z}(1-r^2)))\psi^{L}(\emph{x}_{3}, \emph{b}_{3})+4r r_{c}\overline{\eta}\psi^{t}(\emph{x}_{3}, \emph{b}_{3}))\phi_{\sigma}\\
& -(r^2\emph{x}_{B}+\emph{z}\overline{\eta}(1-r^2))\phi_{s}\psi^{L}(\emph{x}_{3}, \emph{b}_{3})]]\\
& \times \alpha_{s}(t_{d})\emph{C}_{2}(t_{d})h_{d}(\emph{x}_{B}, \emph{z}, \emph{x}_{3}, \emph{b}_{B}, \emph{b}_{3})\exp[-S_{B^{0}_{s}}(t_{d})-S_{M}(t_{d})-S_{\psi}(t_{d})]\},
\end{split}
\end{equation}
\begin{equation}
\begin{split}
{\emph{M}}^{'(V-A)(V-A)}={\emph{M}}^{(V-A)(V-A)}|_{\emph{C}_{2}\rightarrow \emph{a}_{4}},
\end{split}
\end{equation}
\begin{equation}
\begin{split}
{\emph{M}}^{(S-P)(S+P)}=&\frac{32\pi\emph{C}_{\emph{F}}M^{4}_{B^{0}_{s}}}{\sqrt{2\emph{N}_{c}}}\int^{1}_{0}\emph{d}\emph{x}_{B}\emph{d}\emph{z}\emph{d}\emph{x}_{3}\int^\infty_{0}\emph{b}_{B}\emph{b}_{3}\emph{d}\emph{b}_{B}\emph{d}\emph{b}_{3}\phi_{B_{s}}(\emph{x}_{B},b_{B})\\
& \times \{[(\overline{\eta}-r^2)[((1-\emph{x}_{3})(\overline{\eta}+r^2)+\emph{z}(1-r^2)-\emph{x}_{B})\psi^{L}(\emph{x}_{3}, \emph{b}_{3})-r r_{c}\overline{\eta}\psi^{t}(\emph{x}_{3}, \emph{b}_{3})]\phi_{0}\\
& +\sqrt{\eta(1-r^2)}[(\overline{\eta}\emph{z}(r^2-1)-r^2\emph{x}_{B})\phi_{s}\psi^{L}(\emph{x}_{3}, \emph{b}_{3})\\
&+[(\overline{\eta}(\emph{z}(r^2-1)-2r^2(1-\emph{x}_{3}))+r^2\emph{x}_{B})\psi^{L}(\emph{x}_{3}, \emph{b}_{3})+4r r_{c}\overline{\eta}\psi^{t}(\emph{x}_{3}, \emph{b}_{3})]\phi_{\sigma}]]\\
& \times \alpha_{s}(t_{c})\emph{a}_{5}(t_{c})h_{c}(\emph{x}_{B}, \emph{z}, \emph{x}_{3}, \emph{b}_{B}, \emph{b}_{3})\exp[-S_{B^{0}_{s}}(t_{c})-S_{M}(t_{c})-S_{\psi}(t_{c})]\\
& +[(\overline{\eta}-r^2)[((1-r^2)(\emph{x}_{B}-\emph{z}\eta)+\emph{x}_{3}(r^2(\overline{\eta}-\eta)-\overline{\eta}))\psi^{L}(\emph{x}_{3}, \emph{b}_{3})-r r_{c} \overline{\eta}\psi^{t}(\emph{x}_{3}, \emph{b}_{3})]\phi_{0}\\
& +\sqrt{\eta(1-r^2)}[(\emph{z}\overline{\eta}(1-r^2)+r^2\emph{x}_{B})\phi_{s}+(\overline{\eta}(\emph{z}(r^2-1)-2r^2\emph{x}_{3})+r^2\emph{x}_{B})\phi_{\sigma}]\psi^{L}(\emph{x}_{3}, \emph{b}_{3})]\\
& \times \alpha_{s}(t_{d})\emph{a}_{5}(t_{d})h_{d}(\emph{x}_{B}, \emph{z}, \emph{x}_{3}, \emph{b}_{B}, \emph{b}_{3})\exp[-S_{B^{0}_{s}}(t_{d})-S_{M}(t_{d})-S_{\psi}(t_{d})]\},
\end{split}
\end{equation}

with $r_{c}=\frac{m_{c}}{M_{B^{0}_{s}}}$. $C_{F}=\frac{4}{3}$ is the group factor of the $SU(3)_{c}$ gauge group. The $S_{B^{0}_{s}}(t)$, $S_{M}(t)$, $S_{\psi}(t)$ used in the decay amplitudes, the hard functions $h_{i}(i=a, b, c, d)$, and the hard scales $t_{i}$ are collected in the Appendix.

In our work, we also take vertex corrections into account in the factorization diagrams, and the Wilson coefficients are combined in the NDR scheme~\cite{Beneke:1999br,Beneke:2003zv,Beneke:2000ry} as follows:
\begin{equation}
\begin{split}
&\emph{a}_{1}=\emph{C}_{1}+\frac{\emph{C}_{2}}{N_{c}}+\frac{\alpha_{s}}{9\pi}\emph{C}_{2}\times[-18-12\ln(\mu/m_{b})+f_{I}+g_{I}(1-r^2)], \\
&\emph{a}_{2}=\emph{C}_{3}+\frac{\emph{C}_{4}}{N_{c}}+\emph{C}_{9}+\frac{\emph{C}_{10}}{N_{c}}+\frac{\alpha_{s}}{9\pi}\times(\emph{C}_{4}+\emph{C}_{10})
\times [-18-12\ln(\mu/m_{b})+f_{I}+g_{I}(1-r^2)], \\
&\emph{a}_{3}=\emph{C}_{5}+\frac{\emph{C}_{6}}{N_{c}}+\emph{C}_{7}+\frac{\emph{C}_{8}}{N_{c}}+\frac{\alpha_{s}}{9\pi}\times(\emph{C}_{6}+\emph{C}_{8}) [6+12\ln(\mu/m_{b})-f_{I}-g_{I}(1-r^2)],\\
&\emph{a}_{4}=\emph{C}_{4}+\emph{C}_{10}, \\
&\emph{a}_{5}=\emph{C}_{6}+\emph{C}_{8}.
\end{split}
\end{equation}
and the hard scattering functions $f_{I}$ and $g_{I}$ are given in the Ref.~\cite{Cheng:2000kt}, the renormalization scale $\mu$ is chosen at the order of $m_{b}$.

For the $B^{0}_{s} \rightarrow \psi(1D)\pi^+\pi^-$ decay, the amplitude is similar to the decay amplitude of $B^{0}_{s} \rightarrow \psi(2S)\pi^+\pi^-$, just replacing
the DAs of $\psi(2S)$ with the corresponding DAs of $\psi(1D)$ in Eq.~(\ref{eq:amp}).

As for the decay amplitude of the $B^{0}_{s} \rightarrow \psi(3770)(\psi(3686))\pi^+\pi^-$ decay, we give the expression based on the idea of $S-D$ mixing scheme:
\begin{equation}
\begin{split}
{\cal A}(B^{0}_{s}\rightarrow \psi(3770)\pi^{+}\pi^{-})=\cos\theta {\cal A}(B_{s}\rightarrow \psi(1D)\pi^+\pi^-) - \sin\theta {\cal A}(B_{s}\rightarrow \psi(2S)\pi^+\pi^-).
\end{split}
\end{equation}
\begin{equation}
\begin{split}
{\cal A}(B^{0}_{s}\rightarrow \psi(3686)\pi^{+}\pi^{-})=\cos\theta {\cal A}(B_{s}\rightarrow \psi(1D)\pi^+\pi^-) +\sin\theta {\cal A}(B_{s}\rightarrow \psi(2S)\pi^+\pi^-).
\end{split}
\end{equation}

%%%--=================================================================
%%%=====           Numerical evaluation and discussions   ============
%%5===================================================================

\section{Numerical Results And Discussions}\label{sec:numer}

In our numerical calculation, the input parameters are listed in Table \ref{tab}, where the mass of the involved mesons, the lifetime of meson and Wolfenstein parameters are got from 2018 PDG~\cite{Tanabashi:2018oca}. The decay constant of the $\psi(2S)$ is calculated by the leptonic decay process $\psi(2S)\rightarrow e^{+}e^{-}$~\cite{Rui:2016opu}
and the decay constant of the $\psi(1D)$ is calculated in the Ref.~\cite{Wang:2007fs}. The mass of $b$ quark and $c$ quark are running mass which are calculated under the modified minimal substraction scheme at the renormalization scale $\mu$ equals to the quark mass.

\begin{table}[htbp]
\centering
\caption{The input parameters of the $B^{0}_{s} \rightarrow \psi(2S,1D)\pi^+\pi^-$ decay}
\label{tab}
\begin{tabular*}{\columnwidth}{@{\extracolsep{\fill}}lllll@{}}
\hline
\hline
\\
Mass of the involved mesons     &$M_{\psi{(2S)}}=3.686$ {\rm GeV}       &$M_{\psi{(1D)}}=3.77$ {\rm GeV}   &$M_{B^{0}_{s}}=5.367$ {\rm GeV}\\
                                  \\
                                  &$m_{b}=4.2$ {\rm GeV}                   &$m_{c}=1.27$ {\rm GeV}               &$m_{\pi^{\pm}}=0.140$ {\rm GeV}   \\
                                  \\
                                  &$m_{f_{0}(980)}=0.99\pm0.02$ {\rm GeV}            &$m_{f_{0}(1500)}=1.50$ {\rm GeV}  \\
                                  \\
Decay constants                 &$f_{\psi{(2S)}}=296^{+3}_{-2}$ {\rm MeV}                    &$ f_{\psi{(1D)}}=47.8$ {\rm MeV}  \cite{Wang:2007fs}
                                   &$f_{B^0_{s}}=227.2 \pm 3.4$ {\rm MeV} \\
                                   \\
Lifetime of meson               &$\tau_{B^{0}_{s}}=1.509$ {\rm ps}     \\
                                   \\
Wolfenstein parameters           &$\lambda=0.22453\pm0.00044$            &$\emph{A}=0.836\pm0.015$\\
                                   \\
                                   &$\bar{\rho}=0.122^{+0.018}_{-0.017}$   &$\bar{\eta}=0.355^{+0.012}_{-0.011}$\\
                                   \\
\hline
\hline
\end{tabular*}
\end{table}

By using the differential branching ratio formula Eq.~(\ref{equ:dbr}), first we make predictions of branching ratios of decay mode $B^{0}_{s} \rightarrow \psi(2S)\pi^+\pi^-$ for different intermediate state, which including $f_0(980)$ and $f_0(1500)$ two resonances, and the numerical results are listed as follows:

\begin{equation}
{\cal B}(B^{0}_{s} \rightarrow \psi(2S) f_{0}(980) \rightarrow \psi(2S)\pi^+\pi^-)=[7.2^{+1.0+0.1+0.2}_{-0.7-0.0-0.0}]\times10^{-5},
\end{equation}
\begin{equation}
{\cal B}(B^{0}_{s} \rightarrow \psi(2S) f_{0}(1500) \rightarrow \psi(2S)\pi^+\pi^-)=[8.8^{+1.0+0.4+0.4}_{-1.5-0.0-0.1}]\times10^{-7},
\end{equation}

where the three main errors come from the shape parameter $\omega_{B_{s}}$ of the wave function of $B^{0}_{s}$ meson, the hard scale $t$, which varies from $0.9t\sim1.1t$ (not changing $1/b_{i}$, $i=1,2,3$), and the Gegenbauer moment $a_{2}=0.2\pm0.2$~\cite{Aaij:2014emv} in the $\pi\pi$ distribution amplitude, respectively. The other errors from the uncertainty of the input parameters, for example, the decay constants of the $B^0_{s}$ and charmonium mesons and the Wolfenstein parameters, are tiny and can be neglected safely. We can see that the input parameter $\omega_{B_{s}}$ of the $B^{0}_{s}$ meson is the primary source of the uncertainties, which take up approximately $9.7\%\thicksim17.1\%$, and then the Gegenbauer moment and the hard scale $t$, which characterizes the size of the next-leading-order contribution. When we consider the total $S$-wave contributions of the $f_{0}(980)$ and $f_{0}(1500)$, we can get:

\begin{equation}
{\cal B}(B^{0}_{s} \rightarrow \psi(2S)(\pi^+\pi^-)_{S})=[7.6^{+0.9+0.1+0.1}_{-0.6-0.0-0.1}]\times10^{-5},
\end{equation}

which is in agreement with the new experiment data $(7.1\pm1.3)\times 10^{-5}$ in allowed errors~\cite{Tanabashi:2018oca}. Comparing with previous work~\cite{Rui:2017bgg}, we find our calculation of the branching ratio of the $B^{0}_{s} \rightarrow \psi(2S)\pi^+\pi^-$ is more close to the latest experimental results, whose main reason is that we adopt new input parameters in 2018 PDG, and the latest parameters lead to our uncertainties more small.

From the numerical results, we can see that $f_0(980)$ is the principal contribution, which take the percentage of $94.7\%$, just as the experiment observed, and the $f_0(1500)$ is $1.2\%$, while the constructive interference between this two resonance can contribute nearly $4.1\%$ to the total branching ratio.

In experiment, the calculated ratio of the branching fraction have been given in Ref.~\cite{Aaij:2013cpa}, which is

\begin{equation}
\begin{split}
\frac{{\cal B}({B^0_{s}\rightarrow\psi(2S)\pi^+\pi^-})}{{\cal B}({B^0_{s}\rightarrow J/\psi\pi^+\pi^-})}&=0.34\pm0.04({\rm stat})\\
&\pm0.03({\rm syst})\pm0.01({\rm \cal{B}}).
\end{split}
\end{equation}

By using the previous prediction about the branching ratio of the decay mode ${B^0_{s}\rightarrow J/\psi(\pi^+\pi^-)_{S}}$~\cite{Wang:2015uea}, we obtain the ratio ${{\cal B}({B^0_{s}\rightarrow\psi(2S)\pi^+\pi^-})}/{{\cal B}({B^0_{s}\rightarrow J/\psi\pi^+\pi^-})}=0.46^{+0.16}_{-0.18}$, which is consistent with the experiment measurement, and indicate that the harmonic-oscillator wave function for excited charmonium is applicable and reasonable. Besides the decay mode $B^{0}_{s} \rightarrow \psi(2S)\pi^+\pi^-$, we make calculation for the part of $1D$, and we also consider the similar contributions from the containing $S$-wave resonance state, $f_{0}(980)$ and $f_{0}(1500)$, the reason is that these two resonances mass is also within the scope of the $\pi\pi$ invariant mass spectra, which is ${2 m_{\pi}}< \omega <  {M_{B_{s}^{0}}-M_{\psi}}$, after making integral over $\omega$, the results are:
\begin{equation}
{\cal B}(B^{0}_{s} \rightarrow \psi(1D) f_{0}(980) \rightarrow \psi(1D)\pi^+\pi^-)=[1.2^{+0.1+0.0+0.8}_{-0.2-0.1-0.9}]\times10^{-5},
\end{equation}
\begin{equation}
{\cal B}(B^{0}_{s} \rightarrow \psi(1D) f_{0}(1500) \rightarrow \psi(1D)\pi^+\pi^-)=[4.6^{+0.3+0.0+1.5}_{-0.2-0.5-2.1}]\times 10^{-7}.
\end{equation}

Also the total $S$-wave contribution of the ${B^0_{s}\rightarrow\psi(1D)\pi^+\pi^-}$ decay is:
\begin{equation}
{\cal B}(B^{0}_{s} \rightarrow \psi(1D)(\pi^+\pi^-)_{S})=[1.3^{+0.1+0.0+0.8}_{-0.1-0.0-0.6}]\times 10^{-5}.
\end{equation}

 \begin{figure}[htbp]
 \centering
 \begin{tabular}{l}
 \includegraphics[width=0.5\textwidth]{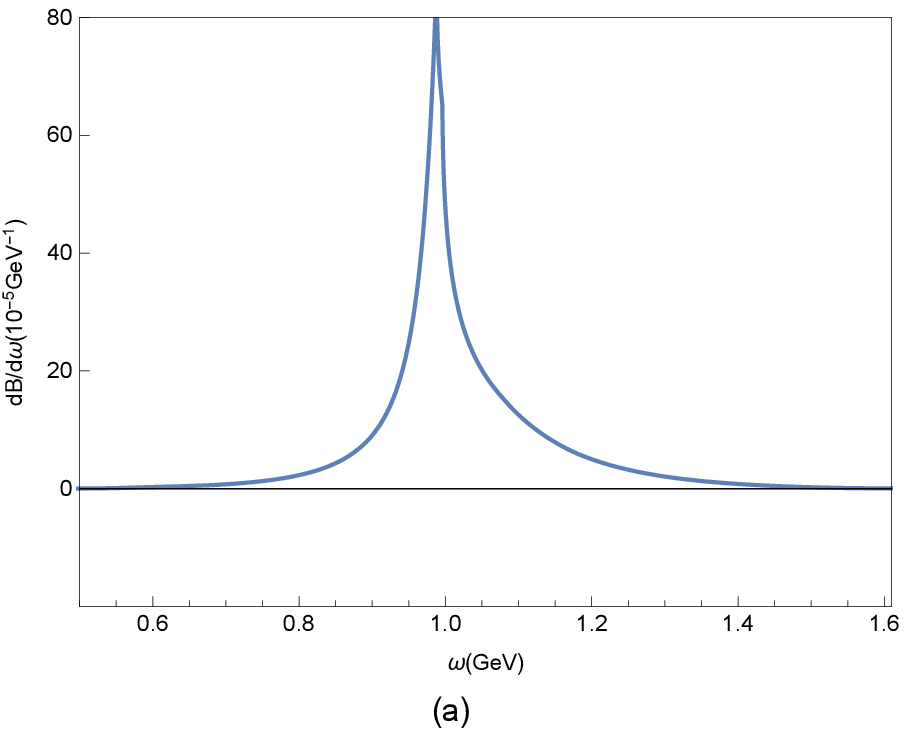}
 \includegraphics[width=0.5\textwidth]{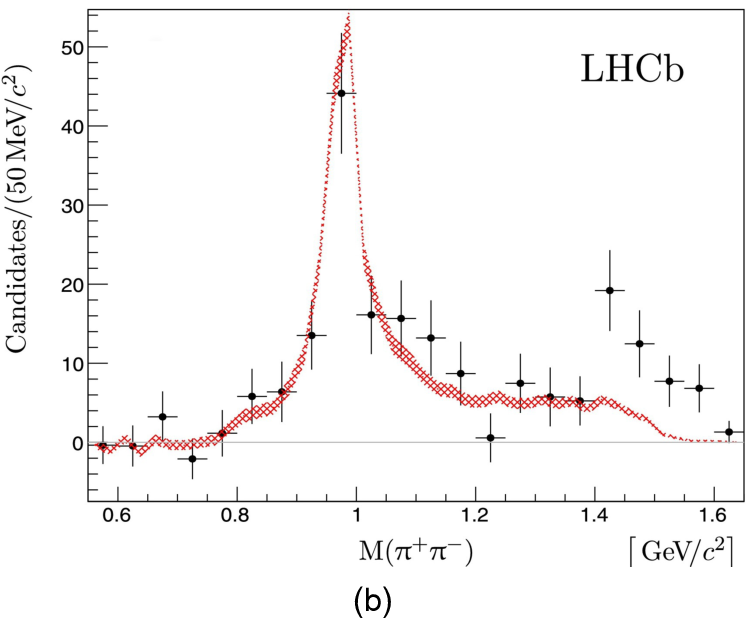}
 \end{tabular}
 \caption {The $S$-wave differential branching ratio of the $B^{0}_{s} \rightarrow \psi(2S)\pi^{+}\pi^{-}$ decay}
   \label{S}
 \end{figure}

 \begin{figure}[htbp]
 \centering
 \begin{tabular}{l}
 \includegraphics[width=0.5\textwidth]{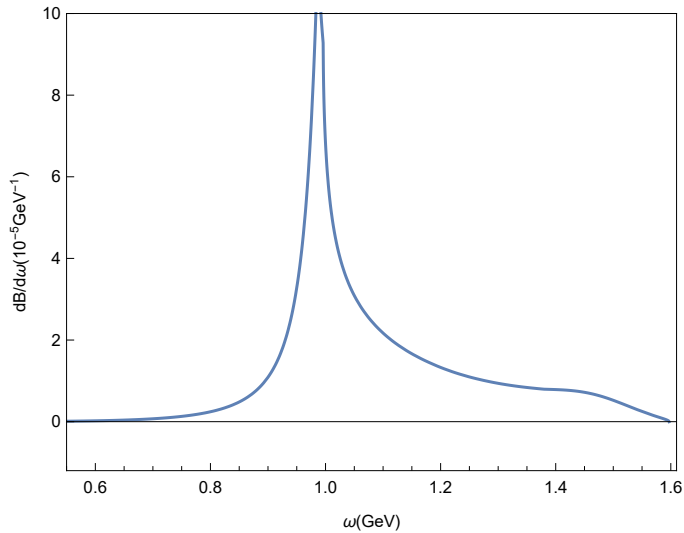}
 \end{tabular}
 \caption {The $S$-wave differential branching ratio of the $B^{0}_{s} \rightarrow \psi(1D)\pi^{+}\pi^{-}$ decay}
   \label{D}
 \end{figure}

In Fig.~\ref{S}(a) and Fig.~\ref{D}, we plot the differential branching ratio of the $B^{0}_{s} \rightarrow \psi(2S,1D)\pi^{+}\pi^{-}$ decay as a function of the $\pi\pi$ invariant mass $\omega$, in which we can clearly see that the peak arises from $f_{0}(980)$, while $f_{0}(1500)$ is unsharp that also make contribution for the decay. For comparison, at the same time, we present the experiment data from LHCb~\cite{Aaij:2013cpa} in Fig.~\ref{S}(b), which shows a basic agreement with our predictive results. Comparing the results between $\psi(2S)$ and $\psi(1D)$, it is easy to find that the results of $\psi(1D)$ is more sensitive to the Gegenbauer moment $a_{2}=0.2\pm0.2$, and this means that although the value is in good agreement with many decay modes, there is still a necessity to explore more accurate data to facilitate a better understanding of the non-perturbative hadron dynamics. In the $\psi(2S)$ and $\psi(1D)$ mode, since $f_{0}(1500)$ mass is near the maximum of $\pi\pi$ invariant mass, the corresponding contributions is very small compared to the total contributions of the $S$-wave. We can note that the branching ratio of the $\psi(1D)$ is smaller than that of the $\psi(2S)$, which should be attributed to the dependence of the corresponding wave function and the decay constant.

Furthermore, we calculate the branching fraction of the mode $B^{0}_{s}\rightarrow \psi(3770)(\psi(3686))\pi^{+}\pi^{-}$ based on $S-D$ mixing scheme, whose two sets of mixing angle has been introduced in Section \ref{sec:intro}, and we list the computational results in Table \ref{3770}.
\begin{table}[htbp]
\centering
\caption{Branching ratios of the quasi-two-decay $B^{0}_{s}\rightarrow \psi(3686, 3770)f_{0}(\rightarrow\pi^ {+}\pi^{-})$ in the pQCD approach based on two sets of $S-D$ mixing angle, where the uncertainties are similar to the previous ones except the last one is the dependence of the mixing angle.}
\label{3770}
\begin{tabular*}{\columnwidth}{@{\extracolsep{\fill}}lllll@{}}
\hline
\hline
                                                      & $\theta=-(12 \pm 2)^{\circ}$       & $\theta=(27\pm2)^{\circ}$   \\
\hline
 \\
$B^{0}_{s}\rightarrow \psi(3686)f_{0}(980)(\rightarrow\pi^ {+}\pi^{-})$  & $ 5.8^{+0.9+0.1+0.5+0.3}_{-0.7-0.0-0.3-0.3}\times10^{-5}$      & $ 8.3^{+1.0+0.0+0.7+0.0}_{-0.8-0.0-0.8-0.1}\times10^{-5}$  \\
\\
$B^{0}_{s}\rightarrow \psi(3686)f_{0}(1500)(\rightarrow\pi^ {+}\pi^{-})$ & $ 6.3^{+0.7+0.3+0.7+0.4}_{-1.0-0.0-0.7-0.5}\times10^{-7}$      & $ 1.3^{+0.1+0.0+0.2+0.0}_{-0.2-0.0-0.1-0.0}\times10^{-6}$  \\
\\
$B^{0}_{s}\rightarrow \psi(3686)(\pi^ {+}\pi^{-})_{S}$                   & $ 6.1^{+0.7+0.0+0.5+0.3}_{-0.6-0.0-0.3-0.3}\times10^{-5}$      & $ 8.8^{+0.9+0.1+0.8+0.1}_{-0.7-0.0-0.8-0.0}\times10^{-5}$           \\
  \\
$B^{0}_{s}\rightarrow \psi(3770)f_{0}(980)(\rightarrow\pi^ {+}\pi^{-})$  & $ 2.7^{+0.2+0.0+0.9+0.2}_{-0.3-0.0-1.1-0.3}\times10^{-5}$      & $ 2.1^{+1.0+0.0+2.2+0.6}_{-0.7-0.1-0.1-0.5}\times10^{-6}$  \\
\\
$B^{0}_{s}\rightarrow \psi(3770)f_{0}(1500)(\rightarrow\pi^ {+}\pi^{-})$ & $ 8.5^{+0.1+0.3+3.5+0.5}_{-0.1-0.0-2.3-0.4}\times10^{-7}$      & $ 1.7^{+0.3+0.0+1.4+0.2}_{-0.9-0.4-0.5-0.1}\times10^{-7}$  \\
\\
$B^{0}_{s}\rightarrow \psi(3770)(\pi^ {+}\pi^{-})_{S}$                   & $ 3.0^{+0.2+0.0+1.1+0.3}_{-0.5-0.0-1.1-0.3}\times10^{-5}$      & $ 2.4^{+1.1+0.0+2.1 +0.6}_{-0.8-0.1-0.4-0.3}\times10^{-6}$           \\
\hline
\hline
\end{tabular*}
\end{table}

Comparing with the pure $D$-wave state, we can notice that the branching ratio of the $S-D$ mixing state for $B^{0}_{s}\rightarrow \psi(3770)\pi^{+}\pi^{-}$ will be raised approximately two times when the mixing angle is $-12^{\circ}$, whose reason is mainly owing to the small decay constant of $\psi(1D)$, which is compatible
with what is summarized in Refs.~\cite{Gao:2006yu,Ding:1991vu,Wang:2007fs,Eichten:1978tg,Heikkila:1983wd}. Moreover, we can observe that the results of the $B^{0}_{s}\rightarrow \psi(3686)\pi^{+}\pi^{-}$ change a little comparing with pure $2S$ mode when taking the mixing effect into account, so $\psi(3686)$ may be regarded as $\psi(2S)$ state. Considering the size of the data collected in LHC-b, we can expect the measurement of this decay mode coming in the near future, that will help us to understand the structure of $\psi(3770)$ and the three body decay mechanism.

\section{Summary} \label{sec:summary}

In this work, we have calculated the contributions from the $S$-wave resonances, $f_{0}(980)$ and $f_{0}(1500)$, to the $B^{0}_{s}\rightarrow \psi(3770)(\psi(3686))\pi^ {+}\pi^{-}$ decay by introducing the $S$-wave $\pi\pi$ distribution amplitudes within the framework of the perturbative QCD approach. Due to the character of $2S-1D$ mixing scheme of $\psi(3770)$, we calculate the branching ratios of $S$-wave and $D$-wave respectively, and the results indicate that the $f_{0}(980)$ is the main contribution of the considered decay, and the differential result of the $\psi(2S)$ mode is in good agreement with the experimental data. We also analyzed the theoretical uncertainties in this paper, and find that the result of $\psi(1D)$ is sensitive to the Gegenbauer coefficient, which we need more accurate data to understand the non-perturbative hadron dynamics. In the end, by introducing the mixing angle $\theta=-12^{\circ}$ and $\theta=27^{\circ}$, we make further calculation of $B^{0}_{s}\rightarrow \psi(3770)(\psi(3686))\pi^ {+}\pi^{-}$, and our calculations show that the branching ratio can be at the order of $10^{-5}$ based on the small mixing angle $\theta=-12^{\circ}$, which will be tested by the running LHC-b experiments.

%%%--=================================================================
%%%=====            Acknowledgements        ==========================
%%5===================================================================

\section*{acknowledgments}

The authors would to thank Dr. Ming-Zhen Zhou for some valuable discussions. This work is supported by the National Natural Science Foundation of China under Grant No.11875226 and No.11047028, and by the Fundamental Research Funds of the Central Universities, Grant Number XDJK2012C040.

%%%--=================================================================
%%%=====            Appendix       ==========================
%%5===================================================================
\section*{Appendix : Formulae For The Calculation Used In The Text} \label{sec:appendix}
%\numberwithin{equation}{section}
\appendix
\renewcommand{\theequation}{{A}.1}
In this section, we list the explicit form of the formulae used above, the Sudakov exponents are defined as:
\begin{eqnarray}
\begin{split}
&S_{B^0_{s}}=s(\emph{x}_{B}p^+_{1},\emph{b}_{B})+\frac{5}{3}\int^{t}_{1/\emph{b}_{B}}\frac{\emph{d}{\bar{\mu}}}{\bar{\mu}}\gamma_{q}(\alpha_{s}({\bar{\mu}})),\\
&S_{M}=s(\emph{z}p^{+},\emph{b})+s(\bar{\emph{z}}p^+,\emph{b})+2\int^{t}_{1/\emph{b}}\frac{\emph{d}{\bar{\mu}}}{\bar{\mu}}\gamma_{q}(\alpha_{s}({\bar{\mu}})),\\
&S_{\psi}=s(\emph{x}_{3}p^{-}_{3},\emph{b}_{3})+s(\bar{\emph{x}}_{3}p^-_{3},\emph{b}_{3})+2\int^{t}_{1/\emph{b}_{3}}\frac{\emph{d}{\bar{\mu}}}{\bar{\mu}}\gamma_{q}(\alpha_{s}({\bar{\mu}})),
\end{split}
\end{eqnarray}

where the Sudakov factor $s(Q,b)$ are resulting from the resummation of double logarithms and can be found in Ref.~\cite{Li:2002mi}, and $\gamma_{q}=-\alpha_{s}/\pi$ is the anomalous dimension of the quark.
The hard scattering kernels function $h_{i}(i=a, b, c, d)$ arises from the Fourier transform of virtual quark and gluon propagators and are written as follows:
\renewcommand{\theequation}{{A}.2}
\begin{equation}
\begin{split}
&h_{a}(\emph{x}_{B}, \emph{z}, \emph{b}_{B}, \emph{b})= K_{0}(M_{B_{s}}\emph{b}_{B}\sqrt{\emph{x}_{B}\emph{z}(1-r^2)}) \times [\theta(\emph{b}-\emph{b}_{B})K_{0}(M_{B_{s}}\emph{b}\sqrt{\emph{z}(1-r^2)})I_{0}(M_{B_{s}}\emph{b}_{B}\sqrt{\emph{z}(1-r^2)})+(\emph{b}_{B}\leftrightarrow \emph{b})],\\
&h_{b}(\emph{x}_{B}, \emph{z}, \emph{b}_{B}, \emph{b})= K_{0}(M_{B_{s}}\emph{b}\sqrt{\emph{x}_{B}\emph{z}(1-r^2)}) \\
&\quad \quad \quad \quad\quad\quad\quad\times \begin{cases}
[\theta(\emph{b}-\emph{b}_{B})K_{0}(M_{B_{s}}\emph{b}\sqrt{\kappa})I_{0}(M_{B_{s}}\emph{b}_{B}\sqrt{\kappa})+(\emph{b}_{B}\leftrightarrow \emph{b})], & {\kappa \geq 0}\\
[\frac{\emph{i}{\pi}}{2} \theta(\emph{b}-\emph{b}_{B})H^{(1)}_{0}(M_{B_{s}}\emph{b}\sqrt{|\kappa|})J_{0}(M_{B_{s}}\emph{b}_{B}\sqrt{|\kappa|})+(\emph{b}_{B}\leftrightarrow \emph{b})], & {\kappa < 0}\\
\end{cases}\\
&h_{c}(\emph{x}_{B}, \emph{z}, \emph{x}_{3}, \emph{b}_{B}, \emph{b}_{3})=[\theta(b_{3}-\emph{b}_{B})K_{0}(M_{B_{s}}\emph{b}_{3}\sqrt{\emph{x}_{B}\emph{z}(1-r^2)})I_{0}(M_{B_{s}}\emph{b}_{B}\sqrt{\emph{x}_{B}\emph{z}(1-r^2)})+(\emph{b}_{B}\leftrightarrow \emph{b}_{3})]  \\
 &\quad \quad \quad \quad\quad\quad\quad\quad\quad\times \begin{cases}
K_{0}(M_{B_{s}}\emph{b}_{3}\sqrt{\beta}), & {\beta \geq 0}\\
\frac{\emph{i}{\pi}}{2} H^{(1)}_{0}(M_{B_{s}}\emph{b}_{3}\sqrt{|\beta|}), & {\beta < 0}\\
\end{cases}\\
&h_{d}(\emph{x}_{B}, \emph{z}, \emph{x}_{3}, \emph{b}_{B}, \emph{b}_{3})=h_{c}(\emph{x}_{B}, \emph{z}, \bar{\emph{x}}_{3}, \emph{b}_{B}, \emph{b}_{3}),
\end{split}
\end{equation}
with the $\kappa=(1-r^2)(\emph{x}_{B}-\eta)$, $\beta=r^2_{c}-(\emph{z}(1-r^2)+r^2\bar{\emph{x}}_{3})(\bar{\eta}\bar{\emph{x}}_{3}-\emph{x}_{B})$, where $J_{0}$ is the Bessel function and $K_{0}$, $I_{0}$ are modified Bessel function with $H^{(1)}_{0}(x)=J_{0}(x)+iY_{0}(x)$.
The threshold resummation factor $S_{t}(x)$ have been parameterized in~\cite{Kurimoto:2001zj}, which is:
\renewcommand{\theequation}{{A}.3}
\begin{equation}
S_{t}(x)=\frac{2^{1+2c}\Gamma(\frac{3}{2}+c)}{\sqrt{\pi}\Gamma(1+c)}[x(1-x)]^{c},
\end{equation}
with the parameter $c=0.04Q^2-0.51Q+1.87$ and $Q^2=M_{B}^2(1-r^2)$~\cite{Li:2009pr}.

For killing the large logarithmic radiative corrections, the hard scale $t_{i}$ in the amplitudes are chosen as
\renewcommand{\theequation}{{A}.4}
\begin{equation}
\begin{split}
&t_{a}=\mathrm{max}\{M_{B_{s}}\sqrt{\emph{z}(1-r^2)}, 1/\emph{b}, 1/\emph{b}_{B}\},\\
&t_{b}=\mathrm{max}\{M_{B_{s}}\sqrt{|\kappa|}, 1/\emph{b}, 1/\emph{b}_{B}\},\\
&t_{c}=\mathrm{max}\{M_{B_{s}}\sqrt{\emph{x}_{B}\emph{z}(1-r^2)}, M_{B_{s}}\sqrt{|\beta|}, 1/\emph{b}_{B}, 1/\emph{b}_{3}\},\\
&t_{d}=t_{c}|_{\emph{x}_{3} \rightarrow \bar{\emph{x}}_{3}}.
\end{split}
\end{equation}

%%%--=================================================================
%%%=====           Reference        ==========================
%%5===================================================================

\end{document}